# *A Pragmatist Understanding of Quantum Mechanics*

*Richard Healey*

**Abstract**

Applications of quantum mechanics have led to many successful predictions and explanations of puzzling phenomena, and we now apply quantum mechanics to gain, process, and communicate information in novel ways. We can understand quantum mechanics by understanding how we have applied it. We should not seek agreement on the nature of the world it represents, because this theory does not itself represent the physical world (though its applications do help us to represent it better).

When applied to a quantum state, quantum mechanics yields probabilities for physical events: both state and probability are objective—not because they represent elements of physical reality, but because each exerts normative authority over the beliefs of anyone who accepts quantum mechanics and applies it relative to a physical situation they may (but need not) occupy. These events may be described by statements (about values of magnitudes) that are meaningful in an appropriate environmental context, and quantum mechanics can help one to say when that is. Measurement creates an appropriate context, so here the Born rule indirectly yields probabilities of measurement outcomes. The quantum state of a system does not "collapse" on measurement: a new state must be assigned relative to a physical situation in which information about the outcome is accessible.

Understood this way, there is no measurement problem, and violation of Bell inequalities does not demonstrate "spooky" non-local action. Quantum field theories have no physical ontology of their own: a quantum field is a mathematical object in a model whose application helps to improve and extend our descriptions of the world in other terms. We cannot realise the scenario of Wigner's friend and its recent extensions: but the data that provide overwhelming evidence for quantum mechanics are objective in the same sense as the relative measurement outcomes described in those scenarios.

**1. Introduction**

The title of this book "How to Understand Quantum Mechanics: 100 Years of Ongoing Interpretation" poses a problem while containing a clue to its solution. The problem is that, 100 years after its birth in 1925, heated controversy continues about how quantum mechanics should be understood, despite widespread agreement with David Mermin's judgement that its unbroken track-record of successful applications have made it "the most useful and powerful theory physicists have ever devised". Immediately after issuing this judgement, Mermin (2012) continued

> Yet today [then] nearly 90 years after its formulation, disagreement about the meaning of the theory is stronger than ever. New interpretations appear every day. None ever disappear.

The is still the situation 13 years later.

Adán Cabello (2017) used the same quote from Mermin to preface his own cartography (in a paper he called "Interpretations of Quantum Theory: A Map of Madness"). He went on to describe this situation as "odd and arguably an obstacle for scientific progress, or at least for a certain kind of scientific progress".

The clue to solving this problem is provided by the key word 'interpretation'. This word has both a broader and a narrower sense. Broadly, to give an interpretation of quantum theory is to say how the theory should be understood: narrowly, an interpretation of quantum theory must say what the physical world is like according to quantum theory. But suppose that quantum theory is to be understood as *not* itself saying what the physical world is like. Then we can come to understand quantum theory even though it *has* no interpretation in the narrower sense. If so, we can seek progress toward an interpretation of quantum mechanics in the broader sense, while dismissing the possibility of scientific progress toward an agreed interpretation of quantum mechanics in the narrower sense.

In what follows I offer an understanding of quantum theory and show how it resolves three outstanding conceptual problems: the measurement problem, the problem of nonlocal action, and the problem of specifying an ontology for a quantum field theory. This is not an interpretation of quantum mechanics in the narrower sense: indeed, so understood, quantum mechanics has no such interpretation. But what is involved in understanding quantum mechanics?

Feynman (1965) claimed that nobody understands quantum mechanics, despite his own understanding, as manifested by his many novel applications of the theory. He went on to tell his readers "what nature behaves like" while cautioning against the misguided attempt to understand in terms of a model representing "how it can be like that". We can come to understand quantum mechanics not by trying to describe a world represented by its mathematical models, but by carefully examining how, and to what ends, these models are applied.

It is common to assume that to apply a mathematical model of a theory like quantum mechanics is to take that model to represent (certain aspects of) the physical world, if only approximately, or after making idealizations and/or abstracting from inessential complications. But a model of quantum theory is not applied by taking it directly to represent what Einstein called "elements of physical reality". To apply a quantum model is not to offer a representation of physical reality: It is to offer sound advice on *how* physical reality may be represented and what to reasonably expect of it when so represented. It follows that we can come to understand quantum mechanics not by trying to describe a world represented by its mathematical models, but by carefully examining how, and to what ends, these models are applied to control how we represent the world in non-quantum terminology, and to guide our beliefs about it as so represented.

Today, quantum mechanics is best thought of not as a single theory, but as a general framework within which many specific theories have been, and continue to be, formulated. Non-relativistic quantum mechanics is one specific theory, and the relativistic quantum field theories of the Standard Model of high-energy physics are others. Although different quantum theories do not have models with the same mathematical structure, all models of a quantum theory contain elements that function in the same general way.

Each model includes a mathematical object (such as a wave-function or density operator) that represents a quantum state of a physical system to which the model may be applied: that state's primary role is then to assign a probability to each of a set of pairwise incompatible and jointly exhaustive events involving this system. To each such event corresponds one or more statements asserting that a magnitude $M$ associated with system $s$ ( an "observable" ) has a value that lies in a set $\Delta$ of real numbers (possibly a unit set): I call this a *magnitude claim* and write it as '$M_s \in \Delta$'. Magnitudes familiar from classical physics include components of position $x$, y, z (in some coordinate system); the corresponding components of momentum $p_x$, $p_y$, $p_z$; kinetic energy $T$; and (local values of) electromagnetic field components. Some quantum theories concern non-classical magnitudes such as photon-number; (local values of) the Higgs field; charm, chirality, and flavour.

A quantum model is applied through the Born Rule, whose legitimate application yields probabilities for some such possible events in an appropriate context. I shall shortly say what makes a context appropriate for a legitimate application of the rule. In a simple quantum model the Born Rule takes the form (for single probabilities),

$$\text{prob}_\rho(M_s \in \Delta) = \text{Tr}(\rho \mathbf{P}_M[\Delta])$$

Here $\rho$ is a density operator representing a quantum state assigned to some system $s$, and the statement '$M_s \in \Delta$' locates the value of magnitude $M$ (on $s$) in Borel set $\Delta$ of real numbers. ($\mathbf{P}_M[\Delta]$ is the relevant projection operator from the spectral family defined by the unique self-adjoint operator $\widehat{M}$ corresponding to magnitude $M$).

Like other objective probabilities, those generated in a legitimate application of the Born rule offer reliable advice to any agent able to understand and follow it. This is advice as to the degree of belief one should have in each of a set of alternative possible events, if one were in a specific physical situation blocking better epistemic access to what actually happens. The spatiotemporal location of a localized agent blocks epistemic access to their future: in a relativistic space-time they can access information transmitted only from their causal past. But a situated agent may face additional physical barriers to informational access, as does Wigner while he remains outside the isolated laboratory of his friend who has agreed to perform a quantum measurement inside it (Wigner, 1962).

An objective probability distribution offers advice tailored to a physical situation that a localized agent might occupy: an agent need not occupy that situation to benefit from this advice, nor choose to heed it even if they do occupy it. This makes Born probabilities (like other objective probabilities) relative not to any actual agent or their opinions (and so not in that sense subjective). Born probabilities are relative not to an agent, but to a physical *agent-situation*. They serve the epistemic and practical needs of any agent who happens to be in that situation: by aligning degrees of belief about possible events to their Born probabilities, that agent can make better decisions about uncertain consequences of their actions. But an agent not in that situation (even a situation one supposes to be located in a world containing no agents) can still use Born probabilities to explain phenomena they can neither control nor observe. Applications of quantum theory extend beyond enhancement of an agent's ability to make prudentially rational decisions about how to act.

Because its primary function is to yield Born probabilities, a quantum state is also objective but relative to a physical agent-situation. A system is not in a quantum state: a quantum state may be correctly assigned to a system relative to a physical agent-situation. A system may be correctly assigned more than one quantum state at once, each relative to a relevantly different agent-situation. One can even describe circumstances in which it is correct to assign a system different *pure* quantum states (representable by distinct wave-functions), each relative to a different agent-situation (Wang *et al*., 2022). These distinct state assignments yield different Born probabilities for the same possible events, but each assignment is correct, relative to the corresponding agent-situation. A statement may truly represent a quantum state or Born probability relative to an agent-situation. But that is not its function, and neither the state nor the probability is or represents what Bell(2004) called a beable of quantum theory—a purported element of physical reality.

According to most textbooks, quantum theory predicts the Born Rule probabilities for our observations of alternative outcomes when we measure a magnitude. But those textbooks don't say what a measurement is, when it occurs, and whether there must be someone observing its outcome. A secondary function of a quantum state is to license the application of the Born Rule to yield probabilities of those mutually exclusive magnitude claims with enough content to be assessable as true or as false. A meaningful magnitude claim does represent the physical world (truly or falsely), but the content of the claim does not derive from what it represents. In accord with Brandom's (1994, 2000) inferentialist pragmatism, a magnitude claim derives its content from the web of reliable inferences that link it to other claims, and ultimately to perception and to action. Since the reliability of these inferential links depends on the physical context with respect to which the claim may be assessed, so does the content of the claim. A single magnitude claim may have rich content relative to some contexts, but little or no content relative to others. Only a magnitude claim with a rich enough content is assessable as true or as false relative to a context.

A secondary function of a quantum state is to offer advice on the content of magnitude claims, and so to license application of the Born rule only to those with a rich enough content to be assigned a degree of belief matching its Born probability. The evolving quantum state of a system and its environment in a model of decoherence may be used to assess the content of a magnitude claim. This model is not used to describe or represent the physical process to which it is applied. Its function is rather to help anyone applying the model to gauge inferential reliability, and thereby content. Each magnitude claim '$M_s=m_j$' has enough content to be assigned a Born probability if the reduced quantum state of $s$ in environment $E$ is robustly diagonal in a "pointer basis" of eigenstates $|m_i>$ of $\hat{M}$, the self-adjoint operator associated with $M$. The Born rule is applicable only to a magnitude claim '$M_s=m_j$' on a system whose interaction with its environment closely meets that condition: I then call the physical context a *decoherence context* (or *decoherence environment*) *for M*. It is legitimate to apply the Born Rule to calculate the probability of a magnitude claim only in, or (more precisely) relative to, such a physical context.

When quantum theory is applied to model naturally-occurring physical interactions, decoherence environments occur naturally, with no agents making quantum measurements. Decoherence in such models is extremely rapid, robust, and practically irreversible. It is then legitimate to apply the Born Rule, whether or not an agent has arranged the conditions in which physical interactions produce an outcome of a measurement of a magnitude with a value relative to that decoherence environment.

The Born Rule may be applied to determine the probabilities of events that are not outcomes of measurements: but it may not be applied to yield probabilities for events in which a pair of incompatible magnitudes (represented by non-commuting operators) both take on precise values, since there are no such events. We can describe a decoherence context relative to which any pairwise compatible set of magnitudes take on values that satisfy the same algebraic relations that obtain among the operators that represent them, and which also hold between their values when the magnitudes are measured together. But no-go results (including Kochen and Specker(1967), Bell(1966), Fine(1982)) show that not all these relations can hold among the values of all (pairwise) compatible sets of magnitudes at once. There is no joint probability distribution over the values of all magnitudes at once that matches the Born probabilities for every compatible set. The Born Rule cannot be understood to apply to possessed values of magnitudes: it applies only to values that magnitudes have as assessed in a decoherence context.

It follows that every outcome of a quantum measurement occurs relative to a context marked by decoherence of a quantum state of system and environment. But that quantum state is itself assigned relative to an agent-situation. So not only a quantum state of a system plus its environment, but also the outcome of a measurement on that system, is relative to an agent-situation. Since the joint quantum state is often the same relative to most relevant agent-situations, this relativity of measurement outcome does not always manifest itself. But later sections will exhibit circumstances in which it has important consequences.

## 2. Why there is no measurement problem

Many people believe that quantum mechanics faces a measurement problem that any acceptable interpretation must solve. The problem has been posed in different ways, and some have located more than one problem with the way measurement is treated in quantum mechanics. The measurement problem arises in its starkest form as an inconsistency between what the theory says and what we observe when we measure its "observables". Leggett (2005, 871) put the problem this way:

> [Most] interpretations of quantum mechanics at the microscopic level do not allow definite outcomes to be realized, whereas at the level of our human consciousness it seems a matter of direct experience that such outcomes occur.

Maudlin (1995, 7) expands on two reasons why one might think that quantum mechanics does not allow a unique, definite outcome of a measurement: the quantum state of a system completely specifies its physical properties; and this state always evolves linearly. It follows that in an interaction between a system in a superposed state and a measuring device, that device ends up in a superposed state in which it fails to have any property capable of representing the unique, definite outcome of the measurement.

No such inconsistency arises if one understands a quantum state to play no role in representing or describing physical properties of a system, but to act instead as a source of sound advice on *how* physical events involving a system may be represented and what to reasonably expect of them when so represented. Understood this way, a quantum state assignment can never imply that a system lacks a physical property: it can at most warn against ascribing a particular physical property to a system, either because any such ascription would lack content in that context, or because one should have zero credence that the system has that property relative to that context. But in a quantum measurement of a magnitude, one can consistently expect some determinate outcome with non-zero Born probability to be indicated by the measurement instrument.

Measurement of a magnitude on a quantum system involves an interaction between that system and an apparatus that results in a directly observable event in a measuring or recording instrument which may be taken to display the measurement outcome. The interaction may be modelled in quantum mechanics by the linear evolution of the quantum state of the system together with the whole apparatus and its environment that robustly decoheres the quantum state of the measuring or recording instrument in its "pointer basis". It is then meaningful to say that an event has occurred in which the pointer magnitude of that instrument has taken on a value. What that value is may be determined by direct observation of the instrument. If the apparatus has been designed and set up correctly, and the instrument has been appropriately calibrated, one may take this value to indicate the outcome of the measurement of the magnitude on that system.

Three points are worth stressing here. First, quantum theory can neither predict nor explain why one outcome is observed rather than another: all it can do is to specify the probability of each possible outcome. Second, in modelling the measurement process, at no point is it necessary to assume that the linear (unitary) evolution of the quantum state of a system, and/or apparatus, and/or environment is interrupted by a non-linear physical "collapse" of a quantum state onto an eigenstate, of the system, the apparatus, or anything else. Discontinuous change of a quantum state does not model a physical process: it corresponds to adoption of a new model with a new quantum state, reassigned to reflect information that has become newly accessible in a changed agent-situation about what events have occurred. Third, the determinate outcome occurs relative to the specified decoherence environment. There may be other decoherence environments relative to which the interaction between this system and apparatus results in no outcome and does not count as a quantum measurement of this (or any other) magnitude on the system. A quantum measurement has an outcome only relative to an appropriate decoherence environment. One can observe that outcome by consulting the measurement indication of a well-designed and calibrated apparatus without actually exhibiting a quantum model of the decoherence interactions that make an outcome possible in this context.

## 3. Why there is no non-local action

Some (Maudlin, 2014; Goldstein, Norsen, Tausk, & Zanghi, 2011) maintain that experimentally confirmed violations of Bell inequalities show that the world is non-local— that an event in space-time region R1 that may be the result of a random event or a free human action influences what happens in a space-like separated region R2. But when correctly understood, quantum theory accounts for observed violations of Bell inequalities with no need to suppose there are any non-local influences.

As a probabilistic theory, quantum mechanics can predict and sometimes explain probabilistic phenomena, not the individual events that manifest them. This is how it explains phenomena associated with violation of Bell inequalities. The statistics of measurement outcomes that manifest these cannot be explained locally as the independent effects of a common cause in their past. But the probabilities of these events follow from application of the Born rule to the entangled quantum state assigned to the pairs of quantum systems they involve on the basis of their causal history. This explanation involves no non-local or retro-causal influence. Quantum mechanics does not show that the world is non-local.

Bell inequalities constrain a set of probabilities for the outcomes of various quantum measurements, including joint measurements on each of a pair of quantum systems initiated by operations performed at distant locations. An experiment is taken to violate these when statistics of actual measurement outcomes significantly fail to match probabilities constrained in this way (while closely conforming to probabilities predicted by quantum theory). A Bell test therefore compares statistical data with rival probabilistic models of that data, refuting those constrained by Bell inequalities while supporting those predicted by quantum theory.

Bell stated an intuitive locality *principle* he called local causality and formulated what he considered a more precise probabilistic *condition* based on it. He argued that since ordinary quantum mechanics does not meet the condition it conflicts with the principle. He concluded that ordinary quantum mechanics is not a locally causal theory. Bell argued that some (subsequently verified) predictions of ordinary quantum mechanics are locally inexplicable because it is not a locally causal theory.

His strongest argument for that conclusion appears in a late paper "La Nouvelle Cuisine" reprinted in Bell (2004). Here is his

*Local Causality Principle*

The direct causes (and effects) of events are near by, and even the indirect causes (and effects) are no further away than permitted by the velocity of light. (2004, 239)

He reformulates this principle to make it "sufficiently sharp and clean for mathematics" in the following

*Local Causality Condition*

A theory will be said to be locally causal if the probabilities attached to values of local beables in a space-time region 1 are unaltered by specification of values of local beables in a space-like separated region 2, when what happens in the backward light cone of 1 is already sufficiently specified, for example by a full specification of local beables in a space-time region 3 [a thick "slice" that fully closes the backward light cone of region 1 wholly outside the backward light cone of 2]. (p. 240).

By his term 'local beable' Bell refers to a magnitude used in a theory to represent (what Einstein called) an element of physical reality that is located within a bounded region of space-time.

Bell's argument that ordinary quantum mechanics is not a locally causal theory appeals to a thought-experiment involving a photon variant of Bohm's spin version of the EPR state considered in the famous paper by Einstein, Podolsky, and Rosen (1935). Pairs of photons are created and correctly assigned (what is now called) the Bell polarization state $\Phi^+ = 1/\sqrt{2}(HH + VV)$ relative to agent-situations in the future light-cone of their production event. A (perfectly efficient) polariser and detector is set up on either side of the source to detect passage of any vertically polarised photon while blocking any horizontally polarized photon.

Since the Local Causality condition is stated in terms of what Bell calls local beables, it cannot be applied without first specifying what are the local beables in the Gedanken-experimental situation he describes. But quantum mechanics uses no local beables: it acknowledges no "hidden variables" $\lambda$; and the quantum state is not assigned locally, and nor is it a beable since it should not be understood as a magnitude used to represent an element of physical reality. Quantum mechanics does admit local beables used by *other* theories to describe the experimental set-up. These include the axes *a*, *b* at which the polarizers are set (in each case to pass a vertically polarized photon and block a horizontally polarized photon) in regions 1, 2 respectively, immediately before a photon pair is incident on them, as well as any other variables *c* needed to completely describe the experimental set-up in region 3. Bell takes his Local Causality condition to imply

$$Pr(A, B|a, b, c, \lambda) = Pr(A|a, c, \lambda) . Pr(B|b, c, \lambda) \quad \text{(Factorizability)}$$

Before assessing Bell's argument that ordinary quantum mechanics is not a locally causal theory it is important to clarify and restate this Factorizability condition. Note first that $\lambda$ is vacuous for ordinary quantum mechanics and so may be deleted from this equation. *a*, *b*, *c* are exogeneous parameters under experimental control and should not be represented as random variables in this condition. Consequently (Factorizability) as applied to ordinary quantum mechanic should be restated as follows

$$Pr_{a,b,c}(A, B) = Pr_{a,b,c}(A) . Pr_{a,b,c}(B) \quad \text{(Factorizability)*}$$

Applying the definition of conditional probability, this implies

$$Pr_{a,b,c}(A|B) = Pr_{a,b,c}(A) \quad \text{(Probabilistic Independence)}$$

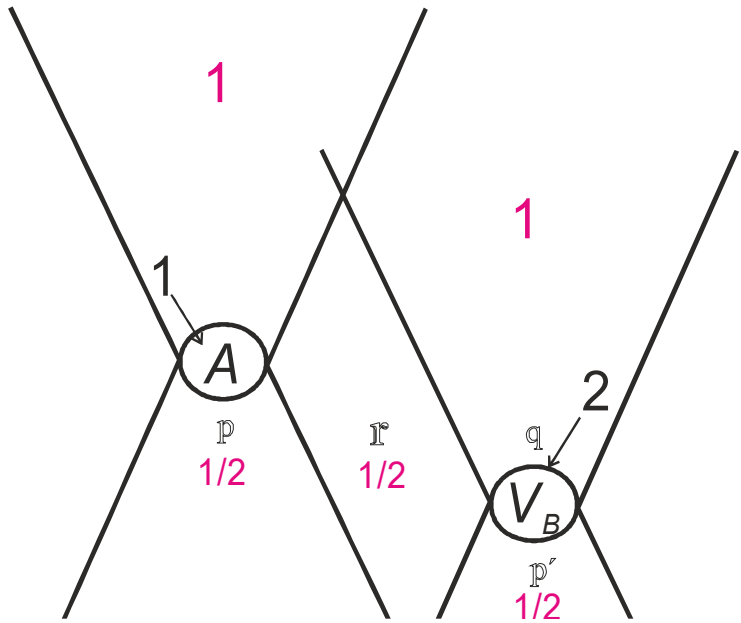

an equation that may be taken to express the probabilistic independence of Alice's result from Bob's result.

> Here is Bell's argument as to why this situation manifests a violation of his condition of Local Causality.
>
> > Each of the counters considered separately has on each repetition of the experiment a 50% chance of saying 'yes'. But when one counter says 'yes' so always does the other, and when one counter says 'no' the other also says 'no', according to quantum mechanics. ... So, specification of the result on one side permits a 100% confident prediction of the previously totally uncertain result on the other side.

But recall that every application of quantum theory's Born Rule yields a probabilistic prediction only relative to an agent-situation, and in this scenario the Born probability is not the same relative to all relevant agent-situations. The following spacetime diagram makes this clear: time variables increase from bottom to top. The numbers 1 or ½ (coloured red online) in five spacetime regions indicate the Born probability, with respect to an agent-situation in that region, of a photon passing Alice's polarizer.

By $p'$ Bob has set the axis of his polarizer to the angle $b$, and by $p$ Alice has set her polarizer's axis to the same angle $a=b$. Relative to an agent-situation such as $q$ in the future light-cone of 2, Bob's counter says 'yes', indicating that the photon on the right was detected as vertically polarized with respect to that axis. The Born probability of Alice's counter saying 'yes' is 1 relative to $q$. But the Born probability of Alice's counter saying 'yes' is ½ relative to $p'$ and also relative to $p$ and to $r$ which are both simultaneous in the lab frame with $q$. Only with respect to an agent-situation such as $q$ in the future light-cone of region 2 could someone use the Born rule to make a 100% confident prediction of the result on Alice's side.

Bell concludes that we have a violation of the Local Causality condition here. But do we? And even if we do, does this imply a violation of the Principle of Locality that Bell used to motivate that condition? To answer these questions, we need to know what are the probabilities quantum mechanics attaches to the values of local beables in region 1. The relevant local beable is the result of Alice's counter, which is either 1 (for 'yes') or 0 (for 'no'). As we saw, quantum mechanics attaches different probabilities to these beables relative to different agent-situations. Relative to $p$, each possible result has probability ½: relative to $q$ the possible results have probabilities 1 and 0 respectively. These probabilities are different, but neither is altered by specification of values of local beables in region 2—since the latter already involves this specification it cannot be altered by simply repeating it. Quantum mechanics does not violate the condition of Local Causality if this is how the condition is correctly applied to it.

This suggests that Bell meant something different by his word 'unaltered'. Consider the probabilities of each possible result at Alice's counter in the scenario depicted in the figure. If the result at Bob's counter is *not* specified, then what is the probability of each possible result of Alice's counter relative to $q$? The Born probability of each of her possible results relative to $q$ gives its objective chance of occurring. Since this depends on the result at Bob's counter, mere failure to specify Bob's result cannot affect that chance. Information as to this result is accessible to any agent located at $q$, so such an agent should use it to become certain that Alice's counter will have the same result.

The objective chance relative to $q$ of Alice's counter giving the same result is 1, in accordance with the Born rule as applied to the correct quantum state |V> of Alice's photon relative to $q$. But an agent at $q$ who lacks, or deliberately sets aside, information as to the result of Bob's counter will consider each possible result of Alice's counter equally credible and assign each probability ½. This is not the objective probability, or chance, relative to $q$ of Alice's counter giving the same result. Accessing information as to Bob's counter's result alters the rational opinion (epistemic probability) of an agent located at $q$ about the result of Alice's counter but does not alter its objective probability. Specification of values of local beables in region 2 may alter an agent's epistemic probabilities for values of local beables in region 1 but it does not alter their objective probabilities.

The last two paragraphs distinguished two ways of understanding probability: as rational opinion based on particular information, and as objective chance as determined by all accessible information. Bell does not say just how he understands probability. But the failure in ordinary quantum mechanics of probabilistic independence as expressed by $Pr_{a,b,c}(A|B) \neq Pr_{a,b,c}(A)$ provides a clue as to his thinking.

Consider data collected in a hypothetical realization of the Gedankenexperiment Bell described. This would almost certainly exhibit relative frequencies matching probabilities that violate the condition $Pr_{a,a,c}(A|B) \neq Pr_{a,a,c}(A)$. Data to be matched against $Pr_{a,a,c}(A)$ would be collected by amassing all instances in which Alice's counter said 'yes' no matter what Bob's counter said. Data to be matched against $Pr_{a,a,c}(A|B)$ would be collected by partitioning those instances into two subsets according to whether Bob's counter said 'yes' or 'no'. Each of the latter subsets will almost certainly yield a relative frequency matching the Born probability of 0 or 1 corresponding to the objective chance relative to $q$ that Alice will get result $A$, as predicted by the Born rule. But the entire set of data to be matched against $Pr_{a,a,c}(A)$ will almost certainly closely match only the *epistemic* probability relative to $q$ of Alice getting result $A$. This corresponds to the epistemic situation of anyone situated at $q$ who knows only that Bob's counter has always produced a result, but not what that result was in each case.

Failure of probabilistic independence here does not show that the objective chance of Alice's result depends on Bob's result. This distinction between objective chance (as specified by the Born rule) and epistemic probability is important in assessing whether a failure of the Local Causality condition would show that the world is non-local. Bell used his intuitive Locality Principle stated in causal terms to motivate his probabilistic Locality Condition. Only if quantum mechanics were to violate the Local Causality condition in such a way that events in one spacetime region may influence events in a space-like separated region could it show that the world is non-local. But events in region 2 at most influence the rational opinion of a hypothetical agent situated at $q$ (or elsewhere in the future light-cone of region 2 where this does not intersect the future light-cone of region 1). They have no influence on what happens in region 1. The objective chance of Alice's result relative to $q$ is not an intrinsic property of Alice's result. Nor does Bob's result influence the objective chance ½ of Alice's result relative to $p$, $r$, or $p'$. The only events that may be influenced by what happens in region 2 lie in the future light-cone of region 2.

Quantum mechanics violates the condition of Local Causality only if that condition is understood in such a way that this violation casts no doubt on the Local Causality principle that Bell used to motivate that condition. Quantum mechanics does not show that the world is non-local.

**4. The ontology of quantum field theories**

Relativistic quantum field theories have been applied very successfully, in the Standard Model of high energy physics and elsewhere. But there is still no agreement on what these theories are about—on the ontology of a quantum field theory. Particles and fields have each been proposed: but philosophers have presented similar reasons for disqualifying both (Fraser, 2008; Baker, 2009). From a pragmatist perspective, this issue may be resolved by acknowledging that it is not the function of a quantum field theoretic model (or any other quantum model) to describe or represent the nature or behaviour of systems to which it is applied.

Mathematical operators such as $\hat{\psi}$, $\hat{\varphi}$ appear in quantum field-theoretic models and these are sometimes called, or taken to represent, physical quantum fields. But that is not the function of these operators when the model is applied, any more than the operators $\hat{x}$, $\hat{p}_x$, represent the magnitudes of the x-position and momentum of a particle in an application of non-relativistic quantum mechanics. One may call the system to which a model is applied a quantum field system, but the application does not represent or describe any of its physical properties. Quantum field theory itself says nothing about these systems or their properties: even to say that a physical system is fermionic or bosonic is not to describe that system, but rather to describe an important feature of the kind of mathematical model successfully applied to it.

A quantum field-theoretic model describes neither fields nor particles. But, when correctly applied, the model yields Born probabilities for alternative possible events described by magnitude statements, and those statements may be about quasi-classical fields or quasi-classical particles. Just as in non-relativistic quantum mechanics, a system's quantum state offers advice on when it is legitimate to apply the Born rule, and to which magnitude statements, through application of a model of quantum decoherence. In some circumstances it is legitimate to apply the Born rule to a quantum field system to yield a probability distribution over a set of mutually incompatible but collectively exhaustive significant magnitude statements about one or more quasi-classical fields: in other circumstances it is legitimate to apply the Born rule to a quantum field system to yield a probability distribution over a set of mutually incompatible but collectively exhaustive significant magnitude statements about one or more quasi-classical particles. The prefix 'quasi' indicates that the field or particle does not possess the full complement of properties ascribed to it by a classical theory, even though it has enough of them to warrant application of the associated term.

Models of the quantum field theory of light (quantized vacuum electromagnetism) have been successfully applied to lasers. The quantum state assigned to a laser may be represented as a vector in Fock space, the symmetrized direct sum of all $n$-dimensional complex Hilbert spaces. An $m$-dimensional subspace is often associated with the presence of $m$ photons of a certain energy in the field, with superpositions of states in subspaces of different dimension associated with an indefinite photon number.

From the present perspective, relative to one kind of decoherence environment in a low powered laser, a statement that the laser light contains $m$ photons, each with a certain energy, is significant enough to be assigned a probability by the Born rule. Relative to a different kind of decoherence environment in a high-powered laser, a statement that the laser light has a frequency in a certain narrow range is significant enough to be assigned a probability by the Born rule (this uses a coherent state decomposition of its Fock space quantum state: "adding a photon of energy $hv$" to a coherent state of frequency $v$ leaves that state unchanged).

An interaction that induces the first type of decoherence environment would permit measurement of the number of photons in the laser light, while an interaction that induces the second type of decoherence environment would permit a measurement of the frequency of that light. In neither case would it be correct to say that its outcome revealed the number of (quasi)-particles or the frequency of the (quasi)-field. In each case, the outcome would be relative to the particular decoherence environment in which it was obtained. A quantum field is a mathematical object in a quantum-field theoretic model that represents nothing present in whatever system that model is applied to. In that sense, a quantum field theory has no ontology.

The 'particle' ontology of non-relativistic quantum mechanics emerges when that theory is thought of as the limit of a quantum field theory. This is the appropriate perspective from which to view issues concerning the identity and individuation of particles in quantum mechanics.

## 5. The scenario of Wigner's friend and its extensions

Wigner (1962) described a thought-experiment and used it to support the view that "it was not possible to formulate the laws of quantum mechanics in a fully consistent way without reference to the consciousness". Though even Wigner himself later came to reject that view, the scenario he described in his thought-experiment continues to play an important role in discussions of the conceptual foundations of quantum theory.

Wigner imagines that in this scenario his friend (a competent and trustworthy physicist) has agreed to perform a particular quantum measurement on a system inside her laboratory after it has been completely physically isolated from all external influences, while Wigner himself remains outside the laboratory. Wigner and his friend agree beforehand that she will prepare the system so that it is correctly assigned a quantum state that is a non-trivial superposition of eigenstates of a particular observable—say the state $1/\sqrt{2}(|\uparrow> - |\downarrow>)$ of a spin ½ particle, whose *z*-spin she is to measure.

His friend performs her measurement and records the outcome: *z*-spin up, or *z*-spin down. The spin-measurement was chosen such that she then correctly reassigns the particle the corresponding eigenstate, $|\uparrow>$ or $|\downarrow>$ respectively. Meanwhile, Wigner correctly assigns some quantum state $\Psi$ to the entire laboratory and all its contents and unitarily evolves that state during the entire period in which his friend was to perform her measurement. After his friend's measurement, neither eigenstate $|\uparrow>$ nor $|\downarrow>$ is compatible with the evolved state $U\Psi$, so Wigner and his friend then disagree about the quantum state of the particle.

Assuming a quantum eigenstate of a system represents it as having the corresponding eigenvalue, they also disagree about whether a measurement has been performed inside the laboratory. If Wigner now enters the laboratory, his quantum state $U\Psi$ predicts that the outcome of his own measurement of the particle's spin will be either *z*-spin up, or *z*-spin down, each with probability ½. But it does not predict that his outcome will match what his friend took to be the outcome of her own measurement. For Wigner there was no measurement outcome until the one he himself obtained after entering the laboratory, which he will find to contain multiple, consonant records of that outcome, including his friend's reported memories. In summary, Wigner and his friend disagree about whether she has performed a measurement with an outcome, what is the resulting quantum state of her lab and its contents, and what is the probability that Wigner will observe a particular outcome on entering the laboratory. Wigner took this conclusion as a *reductio ad absurdum* of the set of assumptions that led to it.

On the present pragmatist understanding, several of these assumptions must be rejected. A system in the isolated laboratory should consistently be assigned distinct quantum states, each relative to one or other of the different agent-situations occupied by Wigner and his friend after she takes herself to have performed her measurement but before he enters her laboratory. These assignments do not yield incompatible descriptions of the system's physical properties, but different but compatible specifications of the objective probabilities of possible outcomes of measurements on it then by one agent or the other, relative to his or her respective agent-situation. Before he enters her laboratory, Wigner is right to deny that his friend has performed a measurement with an outcome relative to his agent-situation, because the laboratory has not interacted with the environment in or outside it in a way correctly modelled by decoherence of its quantum state $U\Psi$ relative to his agent-situation. At the same time, his friend correctly asserts that she has performed her measurement, because relative to her agent-situation her apparatus has interacted with its environment in a way that is correctly modelled by decoherence of its quantum state with respect to its "pointer basis" of eigenstates.

More briefly, Wigner's friend truly states (relative to her agent-situation) that she has performed her pre-arranged measurement with the outcome she reports, while Wigner truly states (relative to his agent-situation) that she has performed no measurement, and so has no outcome to report, despite the fact that he subsequently hears her give such a report and sees the measurement indication and supporting records for himself when he enters her laboratory (thereby changing his agent-situation so that these are all mutually consistent and correct relative to his new agent-situation).

This resolution of "the paradox" of Wigner's friend may seem unnecessarily complex, raising suspicions about the need to relativize a measurement outcome to a decoherence context. As an objection to this resolution, consider the following simpler purported resolution that involves no such relativization.

> Evolution of a quantum state is always unitary but does not represent the dynamical behaviour of a system assigned that state, and so Wigner cannot validly infer from his state assignment $U\Psi$ that his friend has performed no measurement. All he can infer is that, if he were then to observe the laboratory, he would find it just as if his friend had performed a measurement, with outcome "up" or (equally likely) "down". She did indeed perform her measurement whose outcome is recorded in her laboratory as an absolute fact, a fact that Wigner comes to know if he decides to enter her laboratory. But if instead he remains outside the laboratory and has the powers of a super-observer, he could intervene so as to restore the laboratory to its exact pre-measurement condition (as modelled by unitary reversal of Wigner's quantum state $U\Psi \rightarrow \Psi$) thereby erasing all traces of that fact, and indeed of any measurement having been performed in the laboratory.

But recent arguments based on extensions of Wigner's original scenario have cast doubt on this simpler purported resolution by putting more pressure on the idea that the outcome of a quantum measurement is an absolute fact (Bong *et al.* (2020), Haddara and Cavalcanti (2023), Schmid, Ying, and Leifer (2023)). These arguments bear a certain resemblance to earlier no-go results such as those of Bell (1964) and Fine (1982) against understanding the Born rule as assigning probabilities to preexisting values of certain sets of magnitudes. The difference is that these new arguments prove no-go results for probability assignments to the outcomes of measurements of certain sets of magnitudes in situations in which all these measurements are actually performed together.

I will focus on the argument of Haddara and Cavalcanti (2023), an analogue for measurement outcomes of the arguments of Hardy (1993) and Greenberger *et al*. (1990) for preexisting values of observables. In their EWFS observer Charlie is a friend of super-observer Alice, while observer Debbie is a friend of super-observer Bob. Two particles are initially assigned the entangled Hardy state

$$1/\sqrt{3}(|00\rangle + |01\rangle + |10\rangle).$$

In compact spacetime region C, Charlie arranges for his particle to interact with a "probe" system $C$ in a way modelled by the unitary transformation

$$(\alpha|0\rangle + \beta|1\rangle)|r\rangle_C \rightarrow \alpha|0\rangle|0\rangle_C + \beta|1\rangle|1\rangle_C.$$

In a spacelike separated compact region D, Debbie arranges for her particle to interact with a "probe" system $D$ in a way modelled by the unitary transformation

$$(\alpha|0\rangle + \beta|1\rangle)|r\rangle_D \rightarrow \alpha|0\rangle|0\rangle_D + \beta|1\rangle|1\rangle_D.$$

Neither interaction is yet a measurement in the computational basis (0,1), since C and D are not regions of decoherence environments: some call them premeasurements, since measurement further requires environmental decoherence of a pointer observable on a measuring instrument after it has interacted with "probe" system $C$ or $D$ to indicate outcome c or d respectively. What Alice and Bob do, and the consequences of their actions, may be represented in the following spacetime diagram (from Haddara and Cavalcanti (2023): diagonal lines represent light-cones).

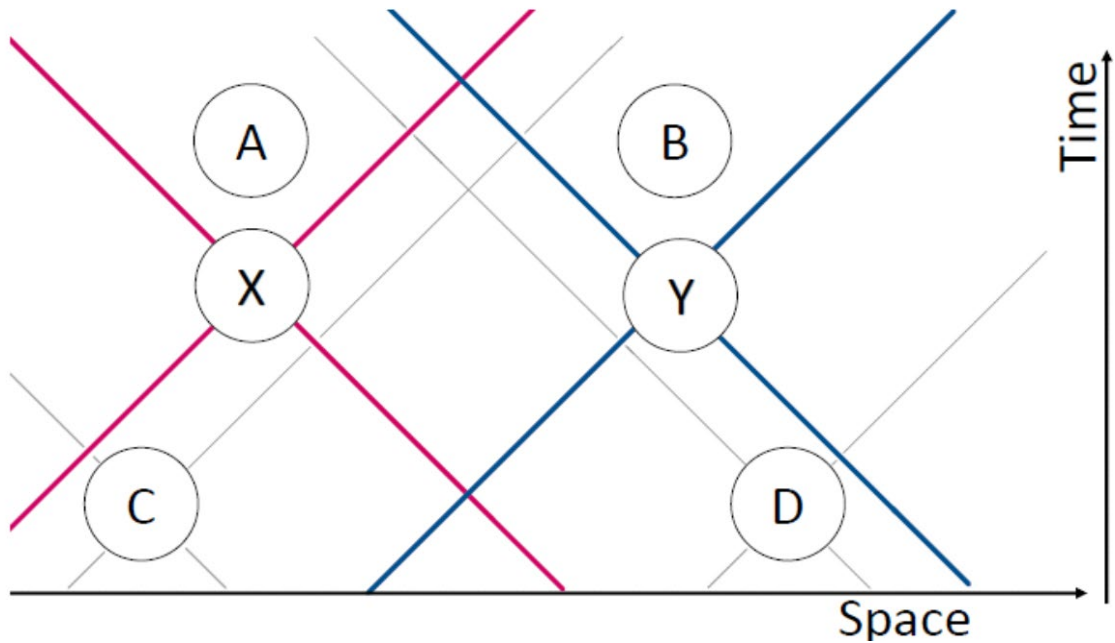

At X, Alice intervenes by freely choosing between different actions to perform at A, each modelled as a unitary transformation:

    i. Measure the pointer observable of Charlie's measuring instrument.

    ii. Close the decoherence environment of Charlie (thereby restoring the entangled Hardy state of the two particles) then measure her own particle in a basis $(+, -)$ incompatible with the computational basis $(0, 1)$.

At Y, Bob makes an analogous free choice of action to perform at B with respect to Debbie's pointer observable and his own particle. The chosen action is implemented by an interaction in region A, B respectively.

    The intended target of the no-go result is the assumption that the observed outcomes of all quantum measurements in this scenario are absolute events, in the following sense:

    *Absoluteness of Observed Events (AOE).* Every observed event is an absolute single event, not relative to anything or anyone.

The argument against AOE of Haddara and Cavalcanti (2023) (following and Bong *et* al. (2020)) makes no assumptions about quantum theory, but it refutes AOE if certain predictions of that theory prove to be correct. Because I am using this no-go result to rebut an objection to a pragmatist understanding of quantum theory that assumes no physical quantum state collapse, I'll take another key assumption of the argument to be that, in the absence of interactions with other quantum systems (including measuring instruments), a quantum state evolves unitarily.

> *Unitary Evolution (U)* The evolution of the quantum state of an isolated system is unitary.

A locality condition weaker than Bell's local causality is placed on the consequences of Alice's and Bob's choices at X and Y.

> *Local Agency (LA)* A super-observer's intervention is uncorrelated with any event outside its future light cone that plays a role in the scenario.

The no-go result is that *LA* and *U* together are incompatible with the absoluteness of events observed by Alice, Bob, Charlie and Debbie assigned extremal (0 or 1) probabilities by application of the Born rule. So, if one accepts *LA*, *U* and the extremal probabilistic predictions of unitary quantum theory in this EWFS scenario, then the outcomes of measurements performed by the super-observers and their friends are not all absolute events. But to reach this conclusion the argument makes a further substantial assumption, which I will discuss later.

> *Tracking* If a super-observer measures the value of the pointer observable indicating the outcome of an observer's measurement, then the value of the super-observer's pointer observable correctly indicates the observer's actual outcome.

To one who understands quantum theory along the pragmatist lines presented here,

this no-go result should not be surprising, since a measurement event occurs only relative to a decoherence environment. To come to understand this or any other EWFS one should begin by saying what decoherence environments it involves. The following figure illustrates the relevant decoherence environments in this scenario.

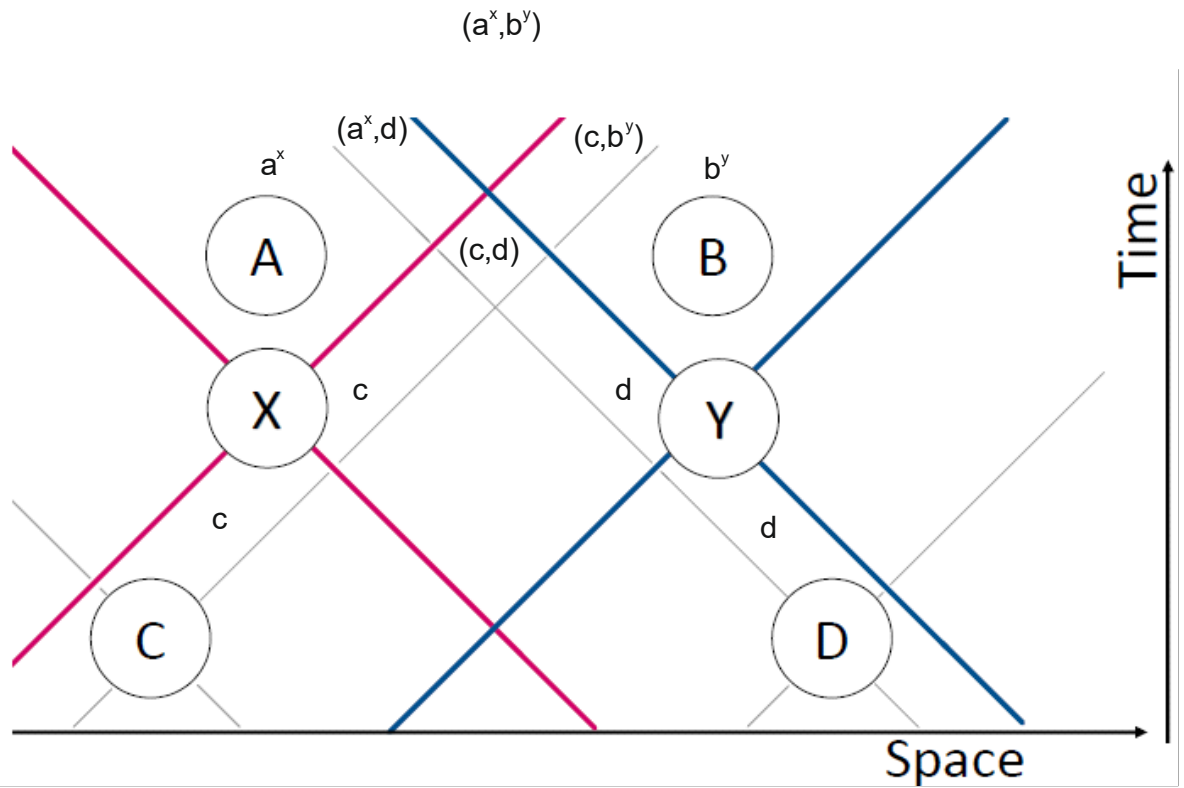

The lower-case letter c in this diagram is placed to roughly indicate the spacetime location of a decoherence environment (DE) in which Charlie's pointer observable *c* on his measuring device indicates the outcome (relative to that DE) of Charlie's measurement in the chosen basis. Other lower-case letters similarly indicate spacetime locations of other DEs corresponding to pointer observables of each observer or super-observer (a for Alice, b for Bob, c for Charlie and d for Debbie). I will now say that the value of a pointer observable of super-observer Alice (relative to a DE corresponding to her measurement of pointer observable *c* of observer Charlie) *tracks* Charlie's outcome if it equals the value of *c* (relative to DE c): analogously, Bob's pointer observable may be said to track Debbie's pointer observable $d$. DE (c,d) is a DE for both c and d. ($a^x$,d) is a DE for both $a^x$ and d; ($b^y$,c) for both $b^y$ and c, and ($a^x$,$b^y$) for both $a^x$ and $b^y$, where the variables (x,y ε {i,ii}) distinguish Alice's and Bob's choices in the following way:

If Alice chooses i, $a^x$ is $a^i$ (a DE for Alice's pointer observable tracking *c*). If Alice chooses ii, $a^x$ is $a^{ii}$ (not a DE for her pointer observable tracking *c*).
If Bob chooses i, $b^y$ is $b^i$ (a DE for Bob's pointer observable tracking *d*). If Bob chooses ii, $b^y = b^{ii}$ (not a DE for his pointer observable tracking *d*).
($a^x$,$b^y$) is a joint decoherence environment for exactly one pair of DEs corresponding to choices (x,y) = (i,i), (i,ii), (ii,i), (ii,ii). Note that only ($a^i$,$b^i$) is a joint DE for the super-observers' pointer readings tracking *c* and *d*.

Given the assumptions *LA*, *U* and *Tracking*, the argument of Haddara and Cavalcanti showed that the extremal probabilities predicted by quantum theory cannot consistently be assigned to absolute outcomes of measurements by all of Alice, Bob, Charlie and Debbie in the EWFS they described. But if these observed events are not absolute, but each occurs (only) relative to a DE in which it is meaningful to speak of it as the outcome of a measurement by one of these four observers, then their occurrence is consistent with the extremal probabilities predicted by quantum theory and with the application of *LA*, *U* and *Tracking*, when suitably understood.

Note that *LA*, *U* and *Tracking* must be suitably understood in their application to this scenario because of the relativization of quantum state assignments to agent-situations and of observed events to decoherence environments. *U* should be understood to apply to a quantum state correctly assigned relative to an agent-situation. *LA* should be understood to apply to an event (such as an observed measurement outcome) that can be said to occur relative to a DE. *Tracking* should be understood as relativized to the respective decoherence environments of the tracking observable and the tracked observable.

Note that interventions at X are correlated only with events in its causal future, and interventions at Y are correlated only with events in its causal future. An intervention either at X or at Y is correlated with events in $(a^x, b^y)$, but not with events in $(c,d)$. This illustrates the importance of taking the outcome of a quantum measurement to be relative not to an observer, but to a decoherence environment arranged (encountered or hypothesized) by an observer.

## 6. Why our measurement outcomes provide objective evidence for quantum theory

A no-go result for absolute outcomes of quantum measurements should not surprise anyone who already understands that an event claimed as the outcome of a quantum measurement occurs only relative to a decoherence context in which that claim may be assessed as both meaningful and true. But it does raise the worry that if that event is not absolute then a claim as to the outcome is or corresponds to a merely relative fact, so that statistics of measurement outcomes can provide no objective data supporting quantum theory even if they match the probabilities give by the Born Rule.

Why do claims reporting the outcomes of quantum measurements constitute objective data for scientists to compare with predictions of quantum theory? A good answer to this question must address two more general issues: What is data, and what makes it objective? Etymology offers unreliable guidance here. Data are not given to scientists: they must be wrested from nature by skilful observation and carefully designed and executed experimentation; and they can tell scientists about events and processes as well as objects.

To support knowledge claims, data must be informative: they need not be expressed in language, but they must have semantic content. As information, an item of data must be embodied in some physical medium (sound waves, ink on paper, state of a computer memory or hard drive, …). An item of data must be capable of distinct embodiments, transferable from one medium to another, preserving and communicating its semantic content (without significant loss). Communication need not be direct and interpersonal, as when two scientists talk to each other: the information may be published in a print or electronic journal, or it may be transferred automatically. But it must be transformable into some physical form that makes its content epistemically accessible to scientists.

Meeting all these conditions already gives data some degree of objectivity, but there are two further conditions met by scientific data that may be thought to limit its objectivity. Data can be informative only about the past. Some philosophers maintain that this is because, since the future does not exist, there is nothing there to be informed about. Less controversially, in a relativistic spacetime, limits on transmission by physical processes exclude receipt of embodied information in a spacetime region that concerns events outside its causal past. Since scientists in spacelike-separated regions can both receive information only from the overlap of their causal pasts, each may have data (temporarily) inaccessible to the other.

This is a situation that scientists rarely need to contemplate, because it quickly resolves itself as events about which they seek information become included in the expanding overlap region of their causal pasts. But it does present itself in loophole-free tests of Bell inequalities such as that by Giustina *et al*. (2015). Their Fig. 2 illustrates data, recorded at each of two different spacelike-separated regions, which could not have been shared among scientists in both regions.

A second restriction on shared access to data arises in EWFSs. In Wigner's original scenario, he lacked access to information about his friend's measurement outcome before entering her laboratory. One could grant him access by *fiat* if one were to assume the *Tracking* condition—a basic epistemological assumption whose rejection in other circumstances would commit one to a form of radical scepticism. But Schmid, Ying, and Leifer (2023, 15) showed how to derive a no-go result for absolute events in a probabilistic generalisation of the argument of (Haddara and Cavalcanti (2023) *without* assuming the *Tracking* condition.

No-go arguments for absolute events do not rule out an understanding of quantum theory according to which a measurement outcome reports an event that occurs relative to a decoherence environment and not absolutely. The tracking condition must now be understood as relativized to appropriate decoherence environments, as follows

> *Relative Tracking* If a super-observer $S$ measures the value of the pointer observable in $DE_O$ indicating the outcome of an observer $O$'s measurement, then the value of the super-observer's pointer observable in $DE_S$ correctly indicates the observer's actual outcome as indicated in $DE_O$.

Like *Tracking*, this relativized assumption is also justified as a basic epistemological assumption whose rejection in other circumstances would commit one to a form of radical scepticism. Once granted, it helps the outcomes, $a^i$ of Alice's or $b^i$ of Bob's measurement outcome, to contribute to the objectivity of the data provided by Charlie's or Debbie's measurement outcomes.

But if Alice or Bob chooses to perform intervention ii, then *Relative Tracking* cannot be applied since this intervention has reset their observer's pointer and undone the decoherence context in which its value had indicated their observer's outcome. Someone may object that it is (timelessly) an objective fact that prior to a super observer's intervention ii that pointer observable did have a value that the intervention cannot alter consistent with *Local Agency*, even though this value may have been subsequently altered or erased. But after an intervention ii, any such fact would no longer have been epistemically accessible and so neither could nor should be considered among objective data for a scientist to compare with predictions of quantum theory. But quantum theory itself should be understood to reject any absolute facts about observers' measurement outcomes, while maintaining that facts relative to a decoherence environment can become epistemically inaccessible although they cannot be altered or erased from spacetime. This situation is illustrated more accurately in Figure 2, where (for example) there is an outcome of Charlie's measurement in DE's marked with the letter 'c', but not in regions marked with a letter $a^{ii}$.

There is a conception of objectivity that may be called *transcendent*. A statement is transcendently objective if and only if it is (or perhaps corresponds to) an absolute fact (what Brukner (2018) called a "fact of the world"). On the other hand, a statement may be said to be *immanently* objective if and only if it is (or corresponds to) a relative fact—a fact relative to (or in) every context of assessment in which the statement is meaningful and true. A decoherence environment for a pointer observable provides a context of assessment for a claim about the outcome of a quantum measurement. No observer need occupy such a context, nor even an agent-situation relative to which it is a decoherence environment. A statement reporting the outcome of a quantum measurement is meaningful only in a decoherence environment in which a magnitude claim about the measuring instrument's pointer observable is meaningful.

A context of assessment for one or more magnitude claims is a decoherence context in which those claims are sufficiently meaningful to be assessed as true or as false. The Born rule yields probabilities only for events in which magnitudes can be meaningfully said to take on values relative to their appropriate decoherence contexts. Such decoherence contexts occur when one or more magnitudes are measured. Statistical summaries of the outcomes of quantum measurements are the immanently objective data scientists treat as the evidence against which they test predictions of quantum theory. When these statistics are judged to match predicted Born probabilities in a wide range of cases while never significantly deviating from them, that is convincing immanently objective evidence warranting acceptance of quantum theory. A metaphysician may yearn for transcendent objectivity, though a pragmatist philosopher will dismiss that goal as unattainable if not unintelligible. Like the rest of science, quantum theory rests on a body of evidence that consists of fallible, but accessible, immanently objective data.

## 7.     Conclusion

We can understand quantum mechanics, but not in terms of some model that represents the physical world to which we apply it. Feynman (1965, 129) warned his audience not to try to understand quantum mechanical processes in terms of some familiar model: but he then went on to describe "what nature behaves like". He did not offer an *unfamiliar* quantum model but rather described what anyone could see for themselves in a suitable experimental arrangement. We can apply quantum mechanics to predict and explain probabilistic features of these phenomena. While itself offering no representation of how nature behaves, quantum mechanics guides and refines the way we describe what is happening, in these and in a host of other familiar as well as unfamiliar circumstances, thereby deepening our understanding of the world.